\documentclass[twoside,12pt]{article}

\usepackage{epsfig}

\newcommand{\be}{\begin{equation}}

\newcommand{\ee}{\end{equation}}

\newcommand{\bea}{\begin{eqnarray}}

\newcommand{\eea}{\end{eqnarray}}

\begin{document}

\title{ \vspace{1cm} Isospin effects in intermediate energy heavy ion collisions}

\author{T.\ Gaitanos$^{1}$, M. Di Toro$^{1}$, G. Ferini$^{1}$, M. Colonna$^{1}$, 
H. H.\ Wolter$^2$
\\
$^1$INFN, Laboratori Nazionali del Sud, Catania, Italy
\\
$^2$Sektion Physik, LMU-M\"unchen, Germany
}

\maketitle

\begin{abstract} 
We investigate the density dependence of the symmetry energy  in a 
relativistic description by decomposing the iso-vector mean field into 
contributions with different Lorentz properties. We find important effects 
of the iso-vector, scalar $\delta$ channel on the density behavior of the 
symmetry energy. Finite nuclei studies show only moderate effects originating 
from the virtual $\delta$ meson. In heavy ion collisions from Fermi to 
relativistic energies up to $1-2~AGeV$ one finds important contributions on the 
dynamics arising from the different treatment of the microscopic Lorentz structure 
of the symmetry energy. We discuss a variety of possible signals which could set 
constraints on the still unknown density dependence of the symmetry energy, when experimental 
data will be available. Examples of such observables are isospin collective flow, 
threshold production of pions and kaons, isospin equilibration and stopping in asymmetric 
systems like $Au+Au$, $Sn+Sn$ and $Ru(Zr)+Zr(Ru)$.
\end{abstract}

\section{Introduction}

Heavy ion collisions at relativistic energies from $0.1$ up to $1-2~AGeV$ offer the 
possibility to access the equation of state (EOS) of nuclear matter under extreme 
conditions of density and temperature \cite{ritter97}. 
Such studies are essential in understanding 
many astrophysical phenomena such as the physical mechanism of supernovae explosions and 
neutron stars. During the last three decades many attempts have been done to investigate 
the properties of highly excited hadronic matter \cite{ritter97}. 

So far asymmetric nuclear matter has been only poorly investigated for densities beyond 
saturation. Finite nuclei studies predict values for the symmetry energy at saturation 
in the order of $30-35~MeV$, however, for supra-normal densities one has to rely on 
extrapolations. On the other hand, in heavy ion collisions highly compressed matter 
can be formed for short time scales, thus the study of such a dynamical process can provide 
useful information on the high density symmetry energy. Recently theoretical studies 
on the high density symmetry energy have been started by investigating heavy ion collisions 
of asymmetric systems \cite{bao,greco} and they have been motivated by 
the planning of new experimental 
heavy ion facilities with neutron rich radioactive beams. 

The aim of this proccedings is to explore the properties of asymmetric nuclear matter within 
a relativistic mean field theory in different nuclear systems, i.e. finite nuclei and 
heavy ion collisions, with the particular interest on understanding the 
high density behavior of the symmetry energy in terms of its Lorentz properties. 

\section{Equation of state of asymmetric nuclear matter}
Within a covariant description of nuclear matter one starts from a Lagrangian of 
an interacting many body system of baryons (protons and neutrons) and mesons which 
characterize the interaction between baryons in terms of baryon-meson vertices. 
In the spirit of a Hartree- or mean field approximation the baryons are given by 
quantum Dirac spinors $\Psi$ and the mesons (iso-scalar, scalar $\sigma$, iso-scalar, vector 
$\omega$, iso-vector, scalar $\delta$ and iso-vector, vector $\rho$) are described by 
classical field equations as follows \cite{liu}
\begin{eqnarray}
& & [\gamma_{\mu}i\partial^{\mu} - g_{\omega}\omega_{0}\gamma^{0} - 
g_{\rho}\gamma^{0}\tau_{3}\rho_{0} - 
(M - g_{\sigma}\sigma - g_{\delta}\tau_{3}\delta_{3})]\Psi = 0
\label{dirac}\\ 
& & m_{\sigma}^2\sigma + B\sigma^2 + C\sigma^3 = 
g_{\sigma}< \hat{\overline\Psi}\hat{\Psi} > 
= g_{\sigma}\rho_s
\label{gordon}\\ 
& & m_{\omega}^2\omega^{\mu} = 
g_{\omega}< \hat{\overline\Psi}\gamma^{\nu}\hat{\Psi} > 
= g_{\omega} j^{\mu}
\label{proca}\\
& & m_{\rho}^2\rho = 
g_{\rho}< \hat{\overline\Psi}\gamma^{0}\tau_{3}\hat{\Psi} > 
= g_{\rho} \rho_{B3}
\label{rhomeson}\\
& & m_{\delta}^2\delta = 
g_{\delta}< \hat{\overline\Psi}\tau_{3}\hat{\Psi} > 
= g_{\delta} \rho_{s3}
\label{deltameson}
\quad .
\end{eqnarray}
In (\ref{gordon}-\ref{deltameson}) the scalar density and the baryonic currents
are given by $\rho_{s},~j^{\mu}=(\rho\gamma,\rho\gamma\beta)$, respectively. 
The corresponding isospin vector and 
scalar densities are then described by $\rho_{B3}=\rho_{p}-\rho_{n}$ and 
$\rho_{s3}=\rho_{sp}-\rho_{sn}$, respectively, with $\rho_{p,n}$ being the proton and 
neutron densities. The iso-scalar, scalar $\sigma$ field contains non-linear 
contributions with parameters $B,~C$. 
The baryon-meson vertices are given by the different 
coupling functions $g_{\sigma,\omega,\rho,\delta}$. Another important quantity is the effective 
Dirac mass which depends on isospin in the presence of the iso-vector, scalar $\delta$ 
meson
\begin{equation}
m^{*}_{i}=M-g_{\sigma}\sigma \pm g_{\delta}\delta
\quad\mbox{(- proton, + neutron)}\quad
\label{effmass}
\quad .
\end{equation}

For the investigation of asymmetric nuclear matter the asymmetry parameter 
$\alpha=\frac{\rho_{n}-\rho_{p}}{\rho_{n}+\rho_{p}}$ is defined which describes the 
relative ratio of the neutron to proton fraction of the nuclear matter. 
The symmetry energy $E_{sym}$ is defined from the expansion of the 
energy per nucleon $E(\rho_{B},\alpha)$ in terms of the asymmetry parameter
\begin{equation}
E_(\rho,\alpha) = E(\rho) + E_{sym}(\rho)\alpha^{2}
+{\cal O}(\alpha^{4})+ \cdots 
\label{esym1}
\end{equation}
with the abbreviation 
\begin{equation}
E_{sym} = \frac{1}{2} 
\frac{\partial^{2}E(\rho,\alpha)}{\partial \alpha^{2}}|_{\alpha=0}
=
\frac{1}{2} \rho
\frac{\partial^{2}\epsilon}{\partial \rho_{B3}^{2}}|_{\rho_{B3}=0}
\label{esym2}
\quad .
\end{equation}
From the the energy momentum tensor one otbains for the symmetry energy 
($f_{i} \equiv (\frac{g_{i}}{m_{i}})^{2},~i=\sigma,~\omega,~\rho,~\delta$) 
\cite{liu}
\begin{equation}
E_{sym} = \frac{1}{6} \frac{k_{F}^{2}}{E_{F}} + 
\frac{1}{2}
\left[ f_{\rho} - f_{\delta}\left( \frac{m^{*}}{E_{F}} \right)^{2}
\right] \rho
\label{esym3}
\quad .
\end{equation}

\nopagebreak
\begin{table}[t]
\begin{center}
\begin{tabular}{|l|c|c|c|c|c|c|c|c|c|}
\hline\hline 
       & $f_{\sigma}$ ($fm^2$)   & $f_{\omega}$ ($fm^2$) & $f_{\rho}$ ($fm^2$)     & $f_{\delta}$ ($fm^2$)     & A & B &  \\ 
\hline\hline
   $NL\rho$          &     10.33        &     5.42     & 0.95 &    0.0      &  0.033     &    -0.0048  \\ 
\hline
   $NL\rho\delta$ &     10.33        &     5.42     & 3.15 &    2.5      &  0.033     &    -0.0048  \\ 
\hline
   NL3                 &     15.73        &     10.53    & 1.34 &    0.0      &  -0.01     &    -0.003   \\ 
\hline\hline
\end{tabular}
\end{center}
\caption{\label{table1} 
Nuclear matter saturation properties in terms of $f_{i}$ and 
$B \equiv \frac{C}{g_{\sigma}^{4}}$ for the RMF models using the $\rho$ and 
both, the $\rho$ and $\delta$ mesons for the characterization of the iso vector mean 
field in comparison with the NL3 model. }
\end{table}
The calculations for symmetric and asymmetric nuclear matter can be done by 
solving self consistently Eqs. (\ref{dirac}-\ref{deltameson},\ref{effmass}). 
The parameters of the model have been fixed to nuclear matter saturation 
properties given in table \ref{table1}. In the following we will focus on the 
iso-vector part of the equation of state in terms of the symmetry energy 
$E_{sym}$ given in Eq. (\ref{esym3}).

From Eq. (\ref{esym3}) it is seen that the introduction of the iso-vector, scalar 
$\delta$ channel influences the density dependence of the symmetry energy: 
in order to reproduce the fixed bulk asymmetry parameter $a_{4}=30.5~MeV$ one 
also has to increase the $\rho$-meson coupling $g_{\rho}$, see table \ref{table1}. 
On the other hand, the Lorentz decomposition of the potential part of $E_{sym}$ 
in terms of a vector $\rho$ and a scalar $\delta$ field affects the density dependence 
of the symmetry energy at high densities due to the suppression of the scalar density 
($\rho_{s} \approx \frac{m^{*}}{E^{*}_{F}}\rho$). This will lead to a stiffer symmetry 
energy at supra-normal densities because of the stronger $\rho$-meson coupling when 
the $\delta$ field is taken into acount in this description.

\begin{figure}[t]
\unitlength1cm
\begin{picture}(8.,8.0)
\epsfig{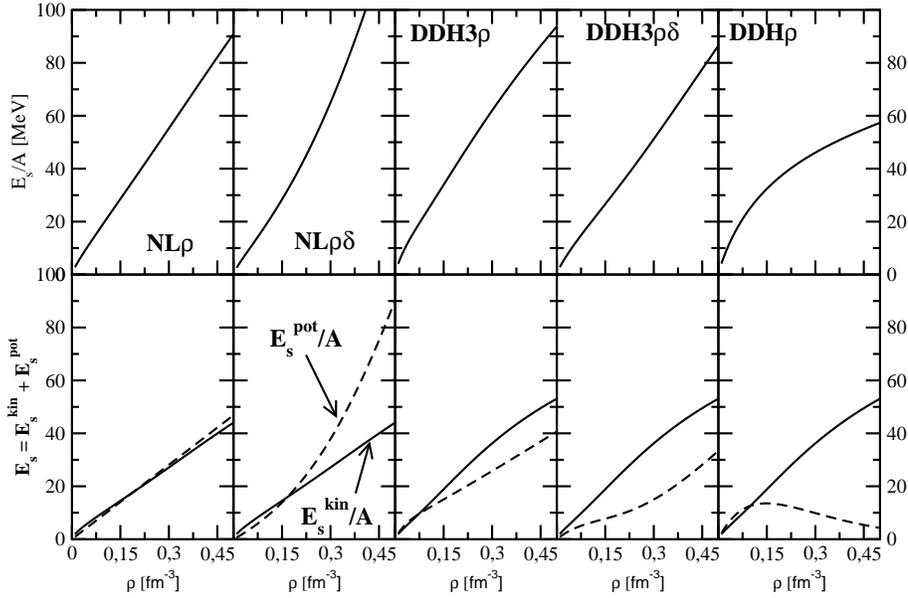}
\end{picture}
\caption{Density dependence of the symmetry energy, Eq. (\protect\ref{esym3}), 
for different models: ($NL\rho, NL\rho\delta$) non-linear Walecka model including 
only the $\rho$-meson and both, the $\rho$ and $\delta$ mesons for the iso-vector 
EOS, respectively. ($DDH\rho, DDH3\rho,DDH\rho\delta$) Same as in the $NL$ cases 
for the iso-vector EOS, but within the Density Dependent Hadronic (DDH) mean field 
theory where all the baryon-meson couplings are explicitely density dependent 
(taken from \protect\cite{gait04}).
}
\label{fig1}
\end{figure}
However, in general the 
situation can be more complicated, i.e. by studying asymmetric nuclear matter within 
more microscopic models such as the Density Dependent Hadronic (DDH) field theory 
\cite{lenske}. In the DDH model the baryon-meson vertices are explicitely 
density dependent with 
a general decrease of the iso-scalar coupling functions ($g_{\sigma,\omega}(\rho)$) 
with respect to the baryon density $\rho$. Such a behavior is consistent with 
realistic Dirac-Brueckner-Hartree-Fock (DBHF) calculations of symmetric nuclear matter 
where no parameters need to be adjusted \cite{jong}. Asymmetric nuclear matter is only 
poorly investigated within the DBHF theory. In Ref. \cite{jong} it was shown that the 
$\rho$ meson coupling strongly decreases with baryon density, but the $\delta$ meson coupling, 
on the other hand, increases for densities above saturation. 

The whole picture is summarized in Fig. \ref{fig1}, where the density dependence of the 
symmetry energy in the spirit of relativistic mean field theory (upper curves) is 
displayed. The bottom panels show separetely the kinetic and potential 
contributions to the total symmetry energy. We used the non-linear Walecka model ($NL$) 
in two different treatments for the iso-vector channel: (a) only with the iso-vector, 
vector $\rho$ meson ($NL\rho$) and (b) with both, the iso-vector, vector $\rho$ and 
iso-vector, scalar $\delta$ mesons ($NL\rho\delta$). The same procedure was applied within 
the DDH theory by fixing the parameters of the iso-vector channel 
($DDH3\rho$ and $DDH3\rho\delta$) to the density dependence of the iso-vector coupling 
functions of the parameter free Dirac-Breuckner model \cite{jong}. 
Finally, in the $DDH\rho$ model, 
which contains only the $\rho$ meson for the description of the iso-vector EOS, the 
parameters were fixed to finite nuclei properties \cite{gait04}. 

We see that the iso-vector, scalar $\delta$ channel has important contributions to the 
symmetry energy for baryon densities above saturation due to the relativistic effects 
as discussed above. However, in the framework of the DDH theory the contribution of the 
$\delta$ meson to the high density symmetry energy is different. Only the comparison between 
$DDH\rho$ and $DDH3\rho\delta$ leads to the same contribution on $E_{sym}$ as the 
corresponding one between $NL\rho$ and $NL\rho\delta$. This is due to the fact that 
in the DDH models the iso-vector couplings has an additional density dependence which 
also contributes to the density behavior of $E_{sym}$, apart from the relativistic 
effects which are always present. 

It is important to realize that the  relativistic effects, i.e. the 
suppresion of the iso-vector, scalar $\delta$ channel for high densities and the 
effective mass splitting between protons and neutrons, lead to a natural momentum 
dependence of the iso-vector EOS even if the baryon-meson vertices do not explicitely 
depend on energy. This important feature is not included in phenomenological 
non-relativistic studies \cite{bao}, where a momentum dependence can be introduced 
in addition, however, with more parameters to be fixed. 

We have applied our models of Fig. \ref{fig1} both to the static case of finite nuclei 
and the dynamical one of heavy ion collisions \cite{gait04}. In finite nuclei only moderate 
effects arising from the $\delta$ meson were found due to the fact that the symmetry 
energy shows a similar density dependence for all the models considered 
for densities at and below saturation. In the next section we thus study the more 
interesting dynamical case where highly compressed asymmetric baryonic matter can 
be formed for short time scales.

\section{Heavy ion collisions at SIS energies: The key observables}
\begin{figure}[t]
\unitlength1cm
\begin{picture}(8.,8.0)
\epsfig{file=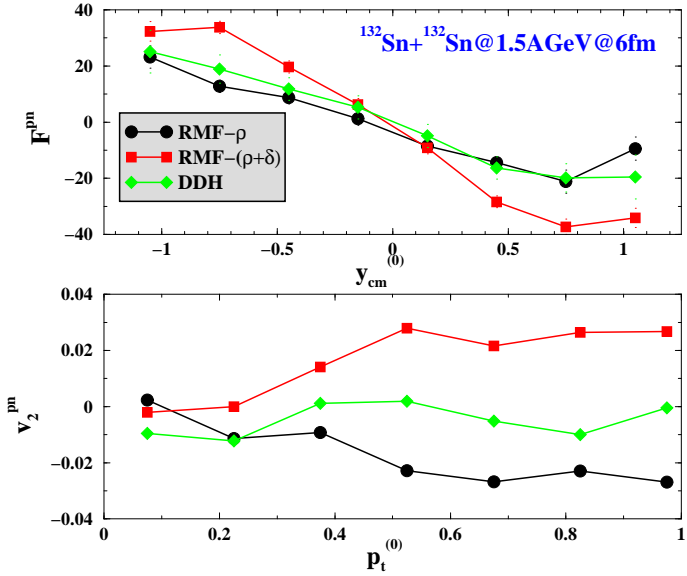,width=9.0cm}
\end{picture}
\caption{(Top) transverse flow $F^{pn}$ as function of the rapidity 
$y_{cm}=\frac{1}{2}\frac{1+\beta_{z}}{1-\beta_{z}}$ ($\beta_{z}$ being the 
component of the velocity along the beam direction) and (bottom) elliptic 
flow $v_{2}^{pn}=\frac{p_{x}^{2}-p_{y}^{2}}{p_{t}}$ as function of the 
normalized transverse momentum $p_{t}^{(0)}=\frac{p_{t}}{(p_{t}^{proj}/A)}$. 
These quantities are calculated from the difference between the proton and 
neutron flows (indicated with the abbreviation $pn$). Calculations with the 
$NL\rho$ (circles), $NL\rho\delta$ (squares) and $DDH\rho$ (diamonds) 
models for a semi-central ($b=6~fm$) $Sn+Sn$ are shown 
(taken from \protect\cite{greco}).
}
\label{fig2}
\end{figure}
In heavy ion collisions (HIC) at SIS energies ($0.1-2~AGeV$) the highly compressed matter 
mainly consists of protons, neutrons and intermediate mass fragments. By choosing 
collisions of asymmetric nucleus like ${}^{197}Au$ or ${}^{132,124,112}Sn$ one can 
hope to {\it see} dense asymmetric nuclear matter at least for some short time 
scales from which one could select sensitive signals related to the symmetry energy at 
supra-normal densities, before expansion and fragmentation sets in. 
The analysis of HIC's with the models discussed in the previous section 
was performed within the relativistic transport equation of a Boltzmann-type (RBUU equation) 
which describes the dynamical evolution of a $1$-particle phase space distribution function 
under the influence of a mean field (depending on the EOS) and binary collisions. 
For a detailed review of the transport theory we refer to Ref. \cite{horror}. 
In the following we discuss some of the most important observables which could set 
constraints on the symmetry energy at high densities. 
{\bf(1) \underline{Collective isospin flows}}\\
An important observable in HIC's is the collective flow due to its high sensitivity on the 
pressure gradients, i.e. on the degree of the stiffness of the EOS at high densities. 
Strong collective flow is related to a more repulsive mean field, i.e. to a stiffer EOS with 
high pressure gradients. There are different components of collective flow: (a) directed 
in-plane flow which describes the dynamics into the reaction plane and can be described 
by the mean transverse in-plane flow $F=<p_{x}(y)>$ as function of the 
rapidity $y$ and (b) elliptic flow which describes the dynamics perpendicularly to 
the reaction plane. The later observable is the most important one due to its earlier 
formation during the high density phase. It can be extracted from a Fourier analysis of 
azimuthal distributions as the second Fourier coefficient $v_{2}$. 

Fig. \ref{fig2} shows the rapidity dependence of the isospin transverse flow (defined as 
the difference in the flow between protons and neutrons) as function of the normalized 
rapidity $y^{(0)}$ and the transverse momentum dependence of the isospin elliptic flow. 
A stronger collective flow is seen with the calculations including the $\delta$ meson 
in the iso-vector channel. This effect becomes very pronounced for the elliptic flow 
$v_{2}$ of  high energetic ($p_{t}^{(0)} \geq ~0.4$) particles due to the fact that 
those particles are emitted earlier during the formation of the high density asymmetric 
matter. 

\begin{figure}[t]
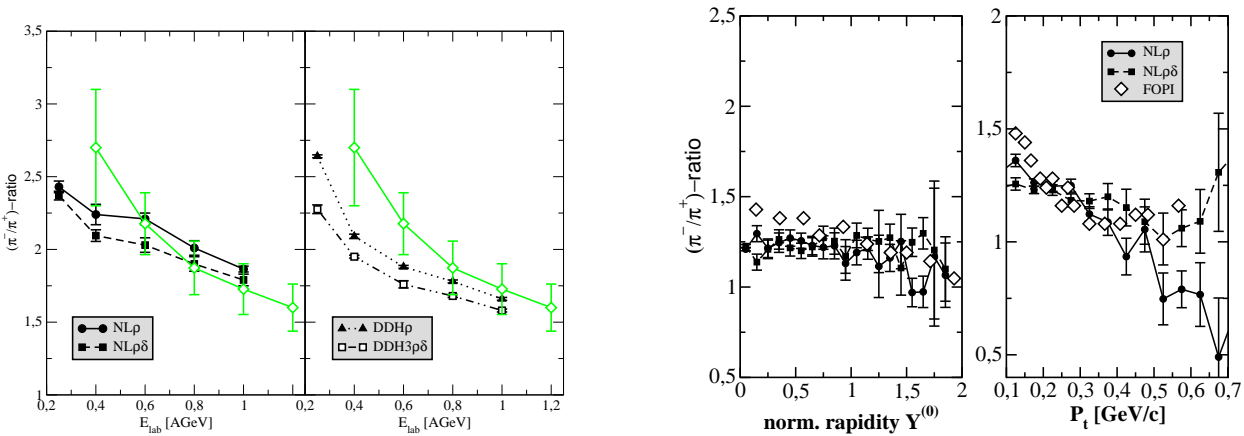

\unitlength1cm
\begin{picture}(8.,8.0)
\put(0.0,0.0){\makebox{\epsfig{file=theo11.12fopi.eps,width=7.5cm}}}
\put(9.,0.0){\makebox{\epsfig{file=theo15.eps,width=7.5cm}}}
\end{picture}
\caption{Left: energy dependence of the $(\pi^{-}/\pi^{+})$-ratio for 
central ($b<2~fm$) $Au+Au$ reactions. Calculations with $NL(\rho,\rho\delta)$ 
and $DDH(\rho,\rho\delta)$ are shown as indicated. Right: 
rapidity ($y^{0}$) and transverse momentum ($p_{t}^{(0)}$) dependence of 
the $(\pi^{-}/\pi^{+})$-ratio for central ($b<1.5~fm$) $Ru+Ru$ reactions with 
$NL\rho$ and $NL\rho\delta$ ($y^{(0)}$ and $p_{t}^{(0)}$ are normalized to the 
corresponding quantities of the projectile per nucleon). The open diamonds shown 
in all the figures are FOPI data taken from \protect\cite{fopi,hong} (the figure 
is taken from \protect\cite{gait04}).}
\label{fig3}
\end{figure}
We can understand the observed effects by referring to Eq. (\ref{esym3}). The 
$\rho$ meson has a repulsive vector character, whereas the $\delta$ meson exhibits 
an attractive scalar one. This Lorentz decomposition is more dominant in the dynamical 
situation due to relativistic effects: the $\rho$ meson linearly increases with the 
Lorentz $\gamma$ factor, whereas the $\delta$ meson is not affected by such dynamical effects 
since the scalar density is a Lorentz scalar quantity. Thus, the stiffness of the symmetry 
energy is effectively enhanced when including the $\delta$ meson in these descriptions 
which yields more repulsion for neutrons than for protons with the net effect of a stronger 
collective dynamics in the $NL\rho\delta$ case.\\
{\bf(2) \underline{Particle production}}\\
Particle production at these high energies is also directly related to the 
dynamics of the earlier high density stage of a heavy ion collision. At SIS 
energies the most dominant inelastic channels are the production of the lowest mass 
resonances $\Delta(1232)$ and $N^{*}(1440)$. The resonances are mainly produced during 
the first nucleon-nucleon collisions and during the high density phase and they 
decay into pions ($\pi^{\pm,0}$). Furthermore, strange particles like 
kaons are created together with hyperons ($Y=\Lambda,\Sigma$) due to 
strangeness conservation through 
baryon-baryon ($BB \longrightarrow BYK^{+}$ with $B=p,n,\Delta$) 
and $\pi$-baryon ($\pi B \longrightarrow YK^{+}$) collisions. 

\begin{figure}[t]
\unitlength1cm
\begin{picture}(8.,9.0)
\epsfig{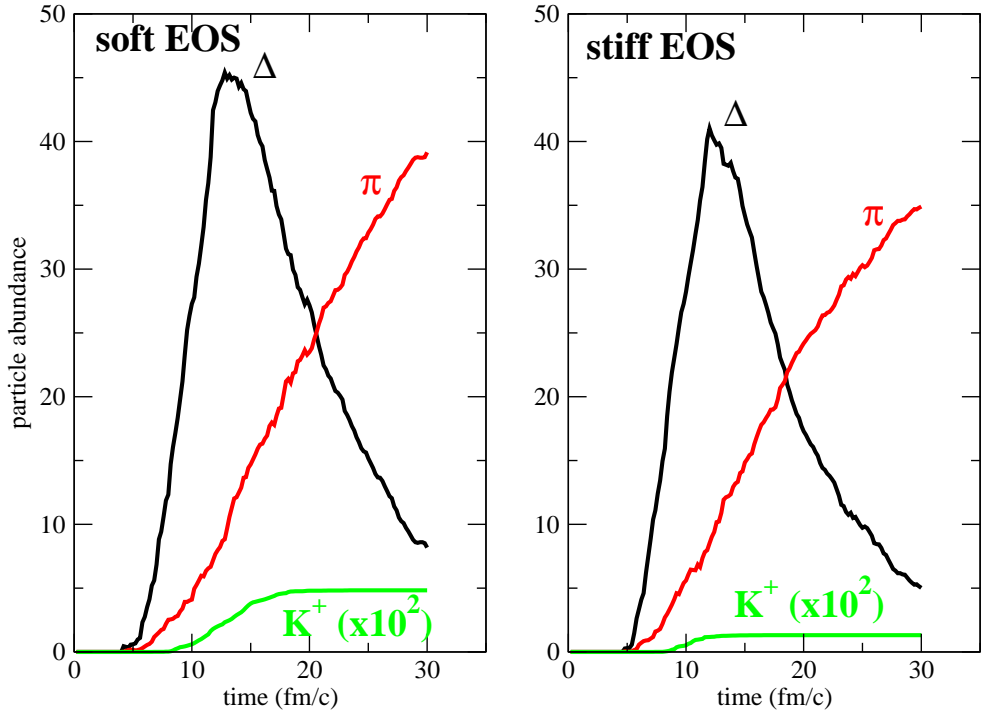}
\end{picture}
\caption{Time evolution of $\Delta$ resonances ($\Delta$), pions ($\pi$) and 
kaons ($K^{+}$) for central ($b=0~fm$) $Au+Au$ reactions at $1~AGeV$ beam energy. 
Calculations with a soft ($NL2$ with a compressibility of $200~MeV$) and stiff 
($NL3$ with a compressibility of $380~MeV$) EOS within the non-linear 
Walecka model are shown (taken from \protect\cite{ferini}).
}
\label{fig5}
\end{figure}
Fig. \ref{fig3} shows the energy, rapidity and transverse momentum dependence of 
the $(\pi^{-}/\pi^{+})$-ratio. This ratio is reduced on the average 
in the models which contain 
the $\delta$ meson in the iso-vector EOS, only for high energetic pions 
($p_{t}^{(0)} \geq 0.35$) the trend turn out to be opposite. The observed isospin effects here 
mainly originate from (a) the different density dependence of the symmetry energy and 
(b) the effective mass splitting ($m^{*}_{n} < m^{*}_{p}$). 

(a) Due to the more stiffer character of $E_{sym}$ neutrons are emitted earlier than 
protons making the high density phase more proton rich. On the other hand, $\pi^{-}$ 
particles are essentially produced via negative charged resonances $\Delta^{-}$, for example 
trhough the process $nn \longrightarrow p\Delta^{-}$, which then decay into 
$\pi^{-}$. Thus due to the earlier neutron emission one observes a reduction of the 
$(\pi^{-}/\pi^{+})$-ratio. This interpretation is also valid for the more complicated 
cases of the $DDH(\rho,\rho\delta)$ models. 

(b) The effective mass splitting leads additionaly to threshold effects since 
in the $(NL,DDH)\rho\delta$ cases less kinetic energy $\sqrt{s^{*}}=m^{*}_{n}+p^{2}$ 
is available for resonance production due to the decrease of $m^{*}_{n}$. 

However, the comparison with very preliminary FOPI data does not yet support any 
definitive conlusion. One reason could be that pions interact strongly with the 
hadronic enviroment due to absorption effects in secondary collisions and the 
Coulomb interaction. Furthermore, with increasing beam energy these secondary effects 
increase (more energy available). 

\begin{figure}[t]
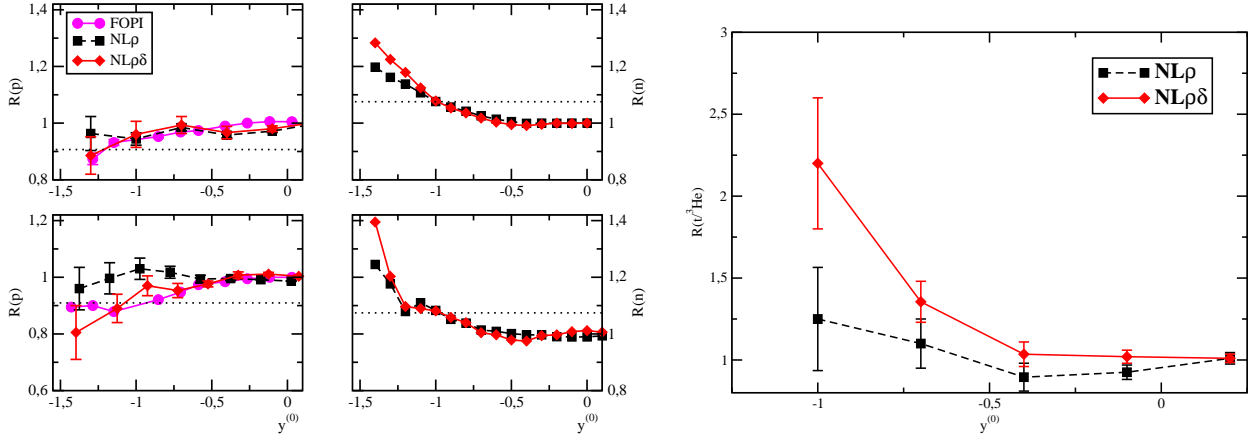

\unitlength1cm
\begin{picture}(8.,8.0)
\put(0.0,0.0){\makebox{\epsfig{file=tracer.all.eps,width=8.5cm}}}
\put(9.,0.0){\makebox{\epsfig{file=TRACERFRA2.E0.4.eps,width=7.5cm}}}
\end{picture}
\caption{Left: rapidity ($y^{(0)}$) dependence of the imbalance ratio 
for protons $R(p)$ (top and bottom on the left) and neutrons $R(n)$ 
(top and bottom on the right) for central ($b=1.5~fm$) mixed reactions 
$Ru(Zr)+Zr(Ru)$ at $0.4~AGeV$ (top) and $1.528~AGeV$ (bottom) beam energy. 
Right: The same but for the ratio of tritons ($t$) to ${}^{3}He$ at $0.4~AGeV$ 
beam energy. Calculations with $NL\rho$ (squares) and $NL\rho\delta$ (circles) 
are shown and compared with FOPI data \protect\cite{hong} as indicated 
(the figure is taken from \protect\cite{gait04b}).
}
\label{fig4}
\end{figure}

We note that pion production takes place over all the collision after compression, 
and thus the differences arising from the high density symmetry energy are not 
very pronounced, however, one want to mention that the density dependence of the 
symmetry energy are not so different between the models used here by comparing with 
other studies, see e.g. \cite{bao}. It might be also more useful to select particles 
directly emitted from the high density region. This can be done by choosing pions 
with high transverse momenta $p_{t}$ \cite{uma}, since in other studies \cite{gait01} it 
was found that baryons are emitted the earlier, the higher their energy or transverse 
momentum is. 
This is seen in Fig. \ref{fig4}, where the differences between $NL\rho$ and 
$NL\rho\delta$ turn out to be more 
important for high transverse momenta $p_{t}^{(0)}$. In particular, for low 
$p_{t}^{(0)} < 0.35$ the $(\pi^{-}/\pi^{+})$-ratio is reduced with the $NL\rho\delta$ 
model, in consistency with the previous discussion. Since the multiplicity is maximal 
at this $p_{t}^{(0)}$ region, on average one obtains a reduction of the 
$(\pi^{-}/\pi^{+})$-ratio with the $NL\rho\delta$ model. However, for high energetic 
particles the situation is different. The reason for the increase of the 
$(\pi^{-}/\pi^{+})$-ratio for $p_{t}^{(0)} >> 0.35$ arises from a combination of 
isospin and Coulomb effects: 
in the earlier stage of the collision the isospin diffusion takes place which populates 
the central shell of the collision with more protons and other positive charged particles. 
Thus, there are two competing effects leading to the observed effect. In one 
hand, one has a population of protons from low to high density regions, but this 
effect is not as strong as compared with the fast neutron emission due to the 
$\delta$ meson. On the other hand, the neutron emission makes the central shell 
more proton-rich and, especially, reduces the production of other negative charged 
resonances. Therefore, the coulomb field acting to high energetic pions, which is 
important at this stage due to high densities, is more repuslive leading the enhancement 
of the pion ratio for pions emitted directly from high density phase space. 
Thus, with the help of coulomb effects one is able to understand the influence of 
the $\delta$ meson on the dynamics of high energetic charged pions and their 
corresponding ratios. 

The kaon production turns out to be a {\it better canditate} for our studies, see 
Fig. \ref{fig5}, since they are produced directly during the high density phase 
without any secondary effects like the pions. The kaon yield strongly depends on the 
EOS, in contrary to the pions, as it can be seen from Fig. \ref{fig5}. Thus, one 
will expect to set more stringent constraints on the high density symmetry energy from 
kaon production since there are a lot of experimental studies. 
Such a progress is under investigation. \\
{\bf(2) \underline{Isospin transparency in the mixed $Ru(Zr)+Zr(Ru)$ systems}}\\
Another interesting aspect in HIC's is the isospin transparency which has been 
extensively studied by experiments of the FOPI collaboration \cite{rami}. 
The idea is to use collisions between equal mass nuclei 
$A=96$, but different isotope ($Ru$ and $Zr$) which can be taken as projectile 
and target by making use of all four combinations 
$Ru(Zr)+Ru(Zr)$ and $Ru(Zr)+Zr(Ru)$. The following imbalance ratio of differential 
rapidity distributions for the mixed reactions $Ru(Zr)+Zr(Ru)$, 
$R(y^{(0)})=N^{RuZr}(y^{(0)})/N^{ZrRu}(y^{(0)})$, 
was considered, where $N^{i}(y^{(0)})$ is the particle yield inside the detector 
acceptance at a given rapidity for $Ru+Zr,~Zr+Ru$ with $i=RuZr,~ZrRu$. The 
observable $R$ can be particularly determined for different particle species, 
like protons, neutrons, light fragments such as $t$ and ${}^{3}He$ and produced 
particles such as pions ($\pi^{0,\pm}$), etc. The observable $R$ chracterizes 
different stopping scenarios. E.g. in the proton case $R(p)$ rises (positive slope) 
for partial transparency, 
falls (negative slope) for full local stopping and is flat when total isospin mixing 
is achieved in the collision. Therefore, $R(p)$ can be regarded as a sensitive 
observable with respect to isospin diffusion, i.e. to properties of the 
symmetry term. 

Fig. \ref{fig4} shows the rapidity dependence of $R$ for different particles and 
energies. With the $NL\rho\delta$ model $R$ decreases for protons and increases for 
neutrons at rapidities near target one. At mid rapidity $R\approx 1$ means full 
isospin mixing, as expected. In an ideal case of full transparency $R$ should 
approach the initial value of $R(p)=Z^{Zr}/Z^{Ru}=40/44=0.91$ and 
$R(n)=N^{Zr}/N^{Ru}=56/52=1.077$ for protons and neutrons at target rapidity, 
respectively. In the calculations one can see that this is approximately the case when 
the $\delta$ meson is taken in the iso-vector channel of the EOS into acount. 
This effect is obvious since in the $NL\rho\delta$ model the neutrons experience 
a more repulsive iso-vector mean field at high densities than the protons and 
it leads to less degree of stopping. This isospin effect is moderate at low, but 
more essential at higher beam energy due to the higher compression in the later 
case. 

It is very important to stress the opposite behavior of $R$ as function of rapidity 
between protons and neutrons which will result to an essential difference between 
$NL\rho$ and $NL\rho\delta$ models for the same observable $R$, in particular, of 
the ratio of $t$ to ${}^{3}He$ fragments. Indeed, it can be seen 
that $R(t/{}^{3}He)$ strongly depends on the consideration of the $\delta$ 
meson in the iso-vector channel, although the corresponding differences for 
$R(p)$ and $R(n)$ are rather moderate for $0.4~AGeV$. Our finding for the 
imbalance ratio $R$ of $R(t/{}^{3}He)$ is in full agreement with a transparency 
scenario which, in particular, becomes more pronounced if the $\delta$ meson 
is taken in these descriptions into acount. Finally, the comparison with FOPI 
data seems to  support a stiffer symmetry energy for high densities, i.e. 
the importance of the $\delta$ meson in the description of asymmetric nuclear matter. 
Corresponding experimental data for $R(t/{}^{3}He)$ would give a more precise 
conclusion. 

\section{Final remarks}
We analyze the relativistic features of the iso-vector part of the equation of state 
by means of a covariant description of symmetric and asymmetric nuclear matter. 
Nuclear matter studies indicate that the stiffness of the symmetry energy is 
mainly dominated by the introduction of a iso-vector, scalar $\delta$ meson which 
significantly changes the Lorentz structure of the iso-vector part of the mean field 
potential at high densities. 

In dynamical situations of heavy ion collisions the high density part of the 
symmetry energy has been studied in terms of different observables which may be directly 
linked to the density dependence of the symmetry energy. Observables which 
are related to the earlier high density phase of the process show the 
strongest effects arising from the different treatment of the iso-vector EOS. 
The collective isospin flow, the transverse momentum dependence of the 
$(\pi^{-}/\pi^{+})$-ratio and the imbalance ratio of clusters seem to be very 
good canditates for studying isospin effects. Also the kaon production might by 
the best observable for such investigations. 

The relativistic decomposition of the iso-vector potential into a vector 
(repulsive $\rho$ field)  and a scalar (attractive $\delta$ field) turns out 
to be essential in understanding the high density behavior of the symmetry energy. 
Furthermore, in dynamical situations of heavy ion collisions the density dependence 
of $E_{sym}$ appears to be effectively stronger affected by the different treatment of the 
iso-vector mean field due to the enhancement of relativistic effects. 
Other important features of a relativistic description are (a) the mean field is 
naturally momentum dependent, important for heavy ion collisions for energies 
above the Fermi one and (b) the effective mass splitting, without introducing any 
additional parameters. 
In this context one should note that in non-relativistic studies the momentum dependence 
of the iso-vector mean field and the splitting in the effective masses between protons 
and neutrons have to be included in addition. 
Finally, the advantage of such relativistic descriptions is a direct comparison with 
more realistic microscopic Dirac-Brueckner-Hartree-Fock theories. 

An this level of investigation we conclude that the symmetry energy should exhibit a 
stiff behavior at supra-normal densities which can be achieved by the introduction of 
an additional degree of freedom (iso-vector, scalar $\delta$ channel). The comparison 
with microscopic DBHF models supports our findings, however, more heavy ion data 
with radioactive beams are needed to make a final definite statement. \\


\begin{thebibliography}{99}
\itemsep -2pt 
\bibitem{ritter97}
W. Reisdorf and H.G. Ritter, 
{\em Annu. Rev. Nucl. Part. Sci.} {\bf 47}, 663 (1997);\\ 
N. Herrmann, J.P. Wessels, T. Wienold, 
{\em Annu. Rev. Nucl. Part. Sci.} {\bf 49}, 581 (1999).
\bibitem{bao}
Bao-An Li, {\em Phys. Rev.} {\bf C67} (2003) 017601.
\bibitem{greco}
V. Greco et al., {\em Phys. Lett.} {\bf B562} (2003) 215.
\bibitem{liu}
B. Liu et al., {\em Phys. Rev.} {\bf C65} (2002) 045201.
\bibitem{lenske}
C.~Fuchs, H.~Lenske, H.H.~Wolter, 
{\em Phys. Rev.} {\bf C52} (1995) 3043.
\bibitem{jong}
F. de Jong, H. Lenske, 
{\em Phys. Rev.} {\bf C57} (1998) 3099.
\bibitem{gait04}
T. Gaitanos et al., {\em Nucl. Phys.} {\bf A732} (2004) 24.
\bibitem{horror}
W.~Botermans, R.~Malfliet, 
{\em Phys. Rep.} {\bf 198} (1990) 115.
\bibitem{fopi}
W. Reisdorf (FOPI collaboration), 
private communication of very preliminary data.
\bibitem{hong}
B. Hong (FOPI collaboration), 
GSI-Report 2002.
\bibitem{uma}
V.S. Uma Maheswari et al., 
{\em Phys. Rev.} {\bf C57} (1998) 922.
\bibitem{gait01}
T. Gaitanos et al., 
Eur. Phys. J. {\bf A12}, 421 (2001).
\bibitem{rami}
W. Reisdorf (FOPI collaboration), 
Acta Phys. Polon. B33 (2002) 107;\\
F. Rami et al. (FOPI collaboration), 
{\em Phys. Rev. Lett.} {\bf 84} (2000) 1120.
\bibitem{ferini}
G. Ferini, T. Gaitanos, M. Di Toro, M. Colonna, in preparation.
\bibitem{gait04b}
T. Gaitanos, M. Di Toro, M. Colonna, H.H. Wolter, 
in preparation.
\end{thebibliography}
\end{document}